\documentclass[iop,10pt]{emulateapj}
\slugcomment{{\sc Accepted to ApJ:} March 30, 2015} 
\usepackage{geometry}                % See geometry.pdf to learn the layout options. There are lots.
\usepackage{graphicx}
\usepackage{amssymb}
\usepackage{epstopdf}
\usepackage{url}

\usepackage[backref,breaklinks,colorlinks,citecolor=blue]{hyperref}
\usepackage[all]{hypcap}

\usepackage{natbib}
\makeatletter
\newlength \figwidth
\if@twocolumn
\else
\fi
\makeatother

\newcommand{\teff}{$T_{\mathrm{eff}}$}
\newcommand{\logg}{$\log g$}
\newcommand{\vmic}{$v_{\mathrm{mic}}$}
\newcommand{\kepler}{\textit{Kepler}}
\newcommand{\dnu}{$\Delta\nu$}
\newcommand{\numax}{$\nu_{\rm max}$}
\newcommand{\outlierDlogg}{-0.2 dex}
\newcommand{\torresOffset}{0.06 dex}
\newcommand{\torresRMS}{0.07 dex} % only 0.055 if we limit to SNR > 100
\newcommand{\seismicOffset}{0.01 dex}
\newcommand{\seismicRMS}{0.05 dex}

\defcitealias{2005ApJS..159..141V}{VF05}

\begin{document}
\title{Accurate Gravities of F, G, and K stars from High Resolution Spectra Without External Constraints}

\author{John M. Brewer, Debra A. Fischer, Sarbani Basu}
	\affil{Department of Astronomy, Yale University}
	\affil{260 Whitney Avenue, New Haven, CT 06511, USA}
	\email{john.brewer@yale.edu}
	\email{debra.fischer@yale.edu}
	\email{sarbani.basu@yale.edu}
\author{Jeff A. Valenti}
	\affil{Space Telescope Science Institute}
	\affil{3700 San Martin Drive, Baltimore, MD 21218, USA}
	\email{valenti@stsci.edu}
\author{Nikolai Piskunov}
	\affil{Uppsala University}
	\affil{Department of Physics and Astronomy, Box 516, 75120 Uppsala, Sweden}
	\email{nikolai.piskunov@physics.uu.se}

%\date{}                                           % Activate to display a given date or no date

\begin{abstract}
We demonstrate a new procedure to derive accurate and precise surface gravities from high resolution spectra without the use of external constraints.  Our analysis utilizes Spectroscopy Made Easy (SME) with robust spectral line constraints and uses an iterative process to mitigate degeneracies in the fitting process.    We adopt an updated radiative transfer code, a new treatment for neutral perturber broadening, a line list with multiple gravity constraints and separate fitting for global stellar properties and abundance determinations.  To investigate the sources of temperature dependent trends in determining \logg\ noted in previous studies, we obtained Keck HIRES spectra of 42 \kepler\  asteroseismic stars. In comparison to asteroseismically determined \logg\ our spectroscopic analysis has a constant offset of \seismicOffset\  with a root mean square (RMS) scatter of \seismicRMS.  We also analyzed 30 spectra which had published surface gravities determined using the $a/R_*$ technique from planetary transits and found a constant offset of \torresOffset\  and RMS scatter of \torresRMS.  The two samples covered effective temperatures between 5000K and 6700K with \logg\ between 3.7 and 4.6.
\end{abstract}


\maketitle

% ========================================================
% Introduction
%
\section{Introduction}\label{sec:introduction}
For many planets, uncertainty in the stellar radius is the largest contributor to uncertainty in planet radius and density, limiting attempts to understand planet formation theory. Ten years ago, the high quality Spectral Properties of Cool Stars (SPOCS) catalog \citep[][hereafter VF05]{2005ApJS..159..141V} provided a uniform analysis which enabled new insights into planet formation \citet{2005ApJ...622.1102F}. However, the authors cautioned that offsets remained with other catalogs and that a temperature dependent bias existed between the spectroscopic and model isochrone gravities for some stars.

More recent analyses have used non spectroscopic ('external') constraints on either \teff\  or \logg\ to remove this trend.  \citet{2012ApJ...757..161T} showed that the Spectroscopy Made Easy (SME) \citep{1996A&AS..118..595V} analysis used in \citetalias{2005ApJS..159..141V} displayed degeneracies between \teff\ and \logg\ and that the surface gravity could be off by up to 0.5 dex in regions of the HR diagram where the Mg b triplet is a poor gravity constraint.  An offset of 0.5 dex in \logg\ resulted in an a corresponding offset of 400K in \teff.  One recently developed external constraint for stars with transiting planets, used in \citet{2012ApJ...757..161T}, is the stellar density derived from the $a/R_*$ ratio \citep{Sozzetti:2007ej}.  Although transiting systems are fantastic laboratories, only 1\% of planetary systems will be favorably aligned  for transits.  Additionally, this method can suffer from inaccuracies due to contamination from the light of nearby companions \citep{2003ApJ...585.1038S} as well as high impact parameter or large eccentricity of the transiting planet \citep{2013ApJ...767..127H}.  Angular diameter measurements of stars are a gold standard for obtaining effective temperatures when the angular resolution of the interferometer can directly measure the angular stellar diameter and accurate distances are known.  This is an important means of validation but is not possible for more distant stars.

\citet{2009ApJ...702..989V} externally constrained SME by using parallax to determine the bolometric luminosity of the star then combined that with derived spectral parameters to interpolate in a grid of stellar models and obtain a constraint on gravity.  Gaia will eventually give us precise distances for millions of stars; however, many of the \kepler\ planet hosts do not have well-measured distances.

To avoid using external constraints, we have searched for the source of the inaccurate gravity determinations and correlated errors in temperature and metallicity.  Most of the gravity information in the line list of \citetalias{2005ApJS..159..141V} is contained in the damping wings of the Mg I b triplet lines, which are sensitive to pressure changes in main sequence stars cooler than about 6200 K.  The visual spectrum is dominated by lines of \ion{Fe}{1} which is the minority species in the Sun and hotter stars.  With few \ion{Fe}{2} lines in the line list the solution is less sensitive to the ionization equilibrium. Our initial investigation showed that although there were minor improvements which could be made by improving the treatment of pressure broadening there was still a large degeneracy between temperature, gravity, and metallicity which could lead to inaccurate results.

The line list of \citetalias{2005ApJS..159..141V} contains $\sim 1000$ atomic lines spanning 170 \AA\  and includes a handful of the stronger molecular lines and 13 \ion{Fe}{2} lines.  The upgrade of the HIRES CCD from one to three detectors extended the wavelength range and we take advantage of this in our new analysis.  Our line list now contains roughly 7500 lines covering more than 350 \AA.  The addition of new temperature and pressure sensitive lines, including 290 \ion{Fe}{2} lines, has helped to break the degeneracies between temperature, gravity, and metallicity. In addition to the large line list, we also use the prescription for broadening by neutral hydrogen from \citet{1998MNRAS.300..863B} when available in VALD-3.  This more accurate treatment of line wing broadening provides a better fit to the wings of strong lines.  Through two independent comparisons, we show that the procedure presented in this paper makes systematic trends in derived surface gravities comparable to or smaller than random errors.

% ============================================================
%	Observations and Reductions
%
\section{Observations and Reductions}\label{sec:observations_reductions}
Our purpose in this work was to improve spectral synthesis modeling for accurate determination of effective temperature and surface gravity.  \kepler\  asteroseismic observations \citep{2010Sci...327..977B} allow precise determinations of stellar mass and radius giving both accurate and precise surface gravities \citep{2011ApJ...730...63G,Basu:2010hv}. We obtained Keck HIRES spectra for a sample of 42 stars covering a range of temperatures, surface gravities, and activity levels to explore how changes in the spectral analysis affected our ability to recover the asteroseismic surface gravity.

\subsection{Spectra}
We obtained 43 spectra of 42 \kepler\  asteroseismic targets using the Keck HIRES spectrograph in the red configuration at a resolution of $R \approx 70,000$. The signal-to-noise ratio (SNR) was typically $> 200$ per pixel column in the region around the Mg I b triplet for all but 4 fainter targets that had S/N of $\approx 120$.  All spectra were reduced using the standard pipeline of the California Planet Search team.  

% ========================================================
% Asteroseismic Surface Gravity
%
\section{Asteroseismic Surface Gravities}\label{sec:astero_logg}
Stellar oscillation frequencies were obtained using short cadence observations from the primary \kepler\  mission \citep{2010Sci...327..977B}.  The short cadence and precision photometry of the \kepler\  telescope allows $\Delta\nu$ and $\nu_{max}$ to be determined extremely accurately for most stars \citep{2014ApJS..210....1C}.

To close approximation the large frequency separation, $\Delta\nu$, scales as $\rho^{1/2}$ where $\rho$ is the mean stellar density and the frequency of maximum power, $\nu_{max}$, scales as $gT_{eff}^{-1/2}$  \citep{2010ApJ...713L.164C}.  Combining these parameters with an externally determined effective temperature, we can use the scaling relations to find the surface gravity.  The uncertainties in \logg\ can be reduced by using a grid of stellar evolutionary models with estimated frequencies instead of simply using the scaling relations. We used the grid-based Yale-Birmingham pipeline \citep{2011ApJ...730...63G,Basu:2010hv} which combines \dnu\ and \numax\ derived from \kepler\  lightcurves with the spectroscopically determined \teff\ and [Fe/H].

\subsection{Iterative Fitting} 
The asteroseismic surface gravities of main sequence stars depend only weakly on the effective temperature and metallicity of the star \citep{2011ApJ...730...63G} with their importance increasing slightly as the star evolves.  We obtained temperature and metallicity from our spectral analysis, then used those values as initial inputs in the asteroseismic analysis.  We then iterated once, fixing the gravity in our spectral analysis to the value returned from asteroseismology and used the newly derived \teff\ and [Fe/H] values in the asteroseismic analysis. The iteration resulted in changes of less than 0.05 dex in the final gravities from those initially determined from the asteroseismic analysis.

% ========================================================
% Spectroscopic Analysis Technique
%
\section{Spectroscopic Analysis Technique}\label{sec:spectroscopic_technique}
SME combines a stellar atmosphere grid, an atomic and molecular line list, and a radiative transfer code to create model spectra based on specified physical parameters such as effective temperature, surface gravity, metallicity, and rotation.  Additionally, SME can fit an observed spectrum using Levenberg-Marquardt least-squares fitting with any number of free global stellar parameters combined with zero or more free elemental abundances.

\citetalias{2005ApJS..159..141V} analyzed nearly 2000 spectra using $\approx 1000$ lines over $\sim 170$\ \AA\ to produce the SPOCS catalog.  Initial line parameters were obtained from the Vienna Atomic Line Database (VALD) \citep{2011BaltA..20..503K}. Then \citetalias{2005ApJS..159..141V} tuned line position as well as $\log(gf)$ and van der Waals broadening coefficients to fit a solar atlas.  In addition, 78 strong molecular MgH and C2 lines from \citep{1993KurCD..18.....K} were also included. Their analysis solves simultaneously for the global parameters surface gravity (\logg), effective temperature (\teff), metallicity ([M/H]), projected rotational velocity ($v \sin i$), and radial velocity ($v_{\rm{rad}}$).  In addition to the global parameters, \citetalias{2005ApJS..159..141V} solved for elemental abundances for Na, Si, Ti, Fe, and Ni.  Microturbulence was fixed at 0.85 km/s and they derived an empirical relation for macroturbulence as a function of effective temperature.  The resulting parameters had good relative precision, though for some stars the spectroscopically determined gravities were inconsistent with isochrone gravities, especially for warmer stars. This later motivated the use of external constraints on surface gravity in a more recent analysis to reduce parameter degeneracy \citep{2009ApJ...702..989V}.

\subsection{Baseline Analysis with New Code}
As a baseline, we began by analyzing the spectra using the line list of \citetalias{2005ApJS..159..141V} with the same version of SME they used.  We then updated the line list over the same wavelength region and updated the version of SME to v439\footnote{Release versions of SME can be downloaded from \url{http://www.stsci.edu/~valenti/sme.html}}, which uses an updated radiative transfer code and updated atmosphere interpolation algorithm.  These changes reduced the differences in spectroscopic gravities relative to our asteroseismic reference values by about half; however, trends in these differences as a function of both metallicity and temperature remained.  Between 5000 K and 6500 K, $\Delta$ \logg\ (spectroscopic minus asteroseismic \logg) spanned 0.5 dex.

We then included additional spectral intervals and decreased the number of simultaneously free parameters in our model to decouple the fitting of global parameters from individual abundances.  These changes dramatically improved the accuracy in our derived \logg\ values, reducing the RMS scatter in $\Delta$ \logg\ to 0.1 dex and removing the trend with respect to \teff.  Further improvements were achieved by iterating the fitting procedure using the derived abundance pattern and allowing the abundances of alpha elements to be free while fitting the global parameters.  This approach removed the trend in derived gravity with respect to [M/H] and reduced the RMS scatter in  $\Delta$ \logg\  to only \seismicRMS.
% t38 was first asteroseismic sample with new line list and SME single step trend reduced to around .2 dex over same teff range.  choice of free alphas finished improvements %

\subsection{Expanded Line List}
The line list now includes more than 350 \AA \ in 20 segments between 5160 \AA \ and 7800 \AA \ and includes nearly 7500 atomic and molecular lines (Table \ref{table:line_segs}).  The lines were obtained from the VALD-3 database \citep{2011BaltA..20..503K} and then tuned to better match a high resolution disk-integrated solar atlas \citep{2011ApJS..195....6W}.  The VALD-3 data contains many astrophysically tuned line parameters,  but we found that about 15\% of the VALD-3 lines still needed adjustment in one or more of line position, $\log(gf)$, or neutral perturber broadening coefficients in order to match the solar spectrum.  Where available, we also used VALD-3 values for the temperature dependent broadening prescription of \citet{1998MNRAS.300..863B}, which gives better fits to the line profiles than the traditional van der Waals formulation.

	\begin{figure}[!h] %  figure placement: here, top, bottom, or page
		\centering
		\includegraphics[width=\columnwidth]{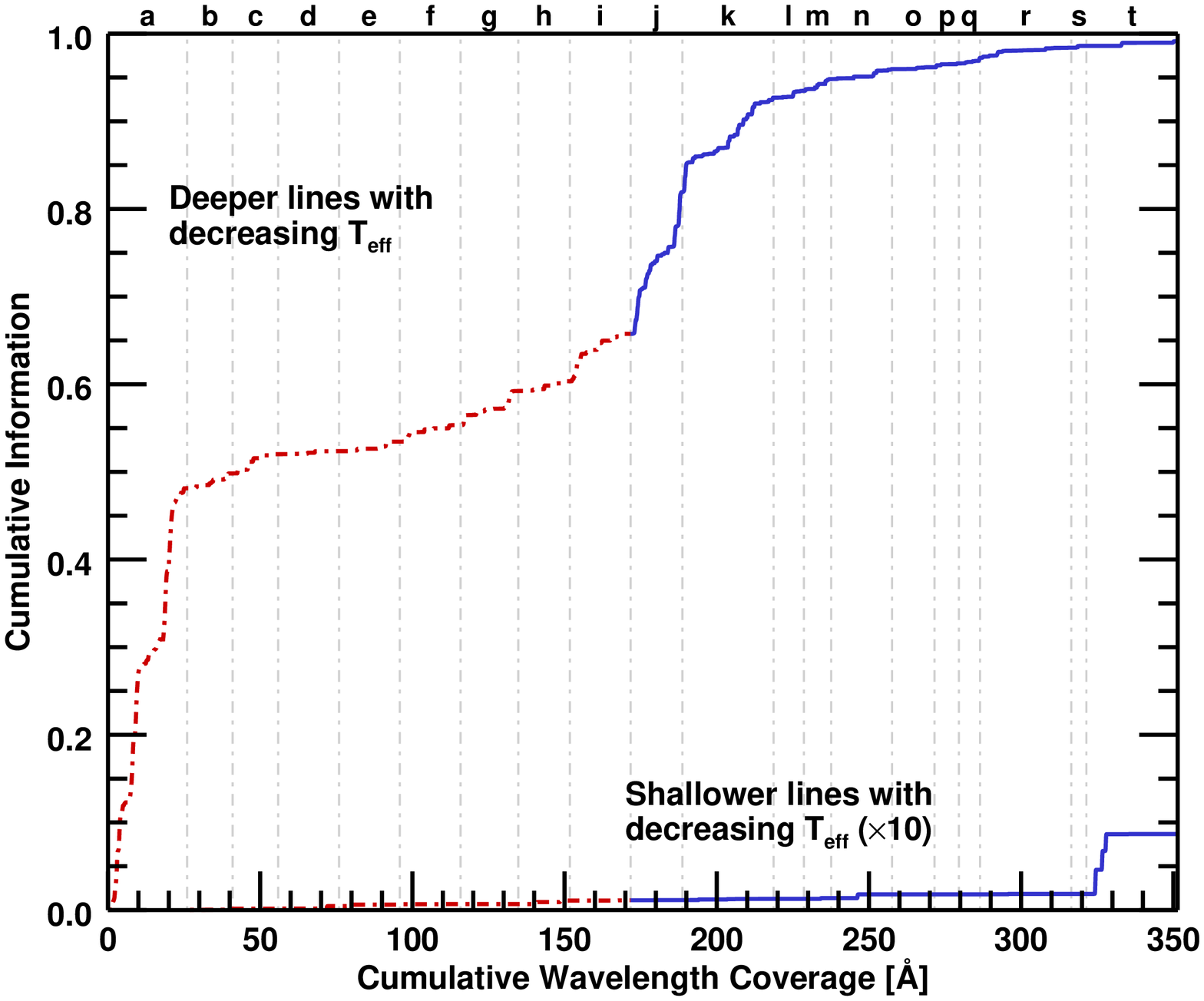}
		\label{fig:teff_info}
		\includegraphics[width=\columnwidth]{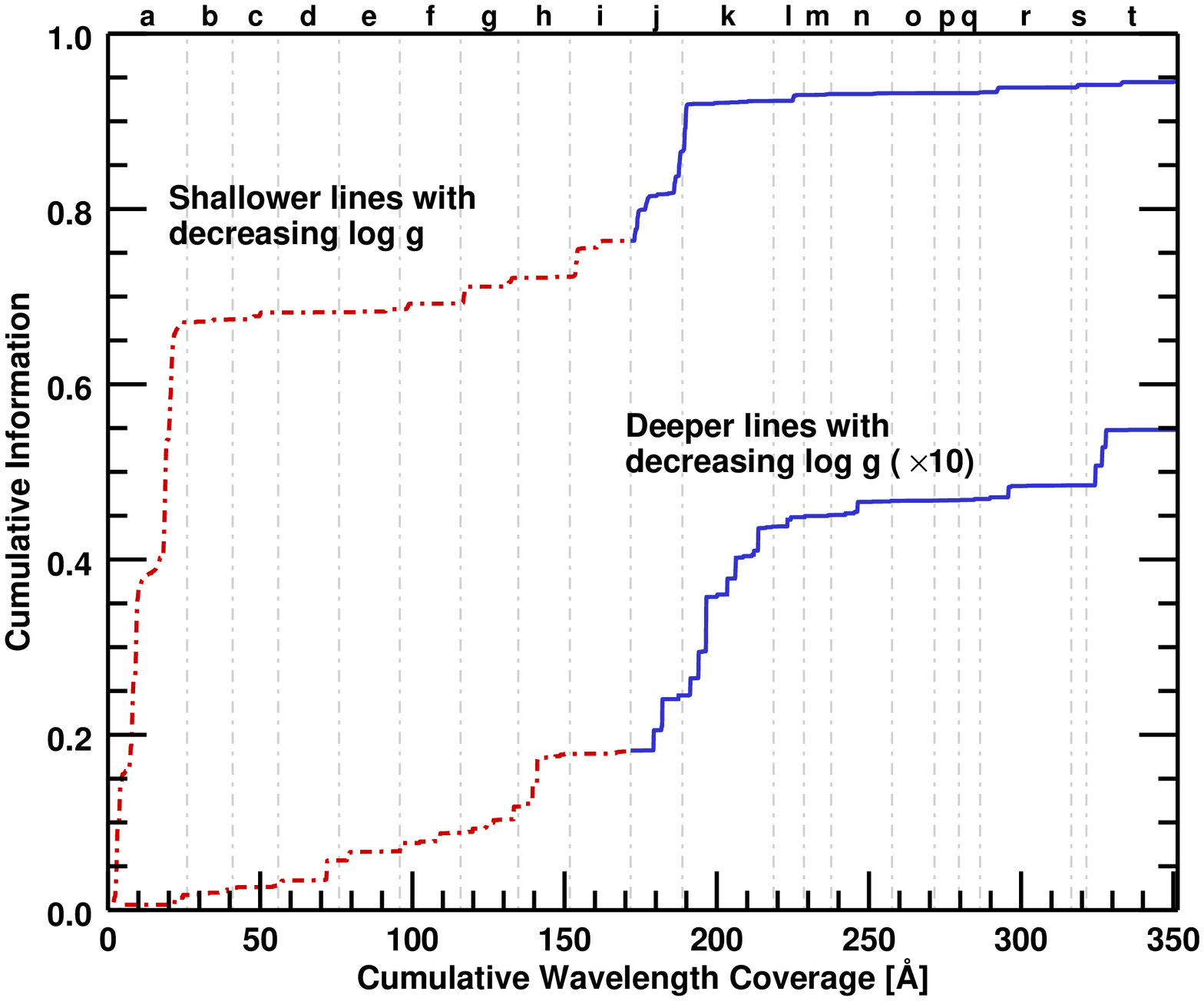}
		\caption{We increased the amount of \teff\ sensitive regions by 51\% and \logg\ sensitive regions by 28\% by including additional wavelength segments. The dotted red line are for those segments which correspond to the original wavelength range of \citetalias{2005ApJS..159..141V}  and the solid blue line corresponds to the newly added wavelength coverage. The gray vertical lines denote the boundaries of the wavelength segments and the letters at the top denote are keys to the wavelength ranges listed in Table \ref{table:line_segs}.}
		   \label{fig:cum_distros}
	\end{figure}
		
We added wavelength regions to increase the number and quality of gravity and temperature dependent lines and to allow abundance determinations for additional elements.  SME calculates $\chi^2$ from the difference between observed and model spectra.  To evaluate the influence that the new segments had on fitting for \logg\ and \teff\ we created a model spectrum with solar parameters, two models with \logg\ adjusted by $\pm 0.2$ dex, and two adjusted by $\pm 200$K in \teff.  We differenced pairs of models and used the squared difference at each unmasked spectrum point as a proxy for the information contributed to $\chi^2$ when fitting to spectra. We then plotted the cumulative distribution of these squared differences to examine this gravity and temperature information as a function of the increasing wavelength point in our unmasked spectrum (Figure \ref{fig:cum_distros}).  The information content of each segment is also detailed in Table \ref{table:line_segs}. The expanded line list added 28\% more gravity information and 51\% more temperature information.  Additionally, although deep gravity sensitive lines tend to get stronger with increasing \logg\  and temperature sensitive lines stronger with decreasing \teff, there are some lines which display the opposite behavior.  As noted by \citet{2008oasp.book.....G}, weak lines of ions or atoms of an element that is predominantly found in the same ionization state (e.g. weak \ion{Fe}{2} lines) provide important gravity information because these lines become stronger with decreasing gravity.  The opposing line growth of these lines is especially helpful in constraining \teff\ and \logg\  and we more than doubled the number of these spectral lines.  Since \ion{Fe}{1} is in the next lower energy state than the dominant \ion{Fe}{2}, weak lines of \ion{Fe}{1} will not be very pressure sensitive and so ionization equilibrium provides an additional gravity constraint to the Mg I b wings, for lines where non-LTE effects do not significantly affect ionization equilibrium.

\begin{table*}
  \centering
  \begin{tabular}{ c l l c c c c c }
\hline 
\hline \\[-1.5ex]
		&			       & 			& \multicolumn{2}{c}{Decreasing \logg} & \multicolumn{2}{c}{Decreasing \teff} &  \\
Segment & $\lambda_{start}$ & $\lambda_{end}$ & lines grow & lines weaken & lines grow & lines weaken & $\mathrm{VF05}$ \\
\\[-1.5ex]
\hline
  a & 5164 &   5190 &   0.17\%   &  67.07\% &  48.14\%  &  0.00\% & x \\
  j & 5190 &   5207 &   0.63\%   &  10.20\% &  16.17\%  &  0.00\% &   \\
  k & 5232 &   5262 &   1.93\%   &   5.77\% &  10.78\%  &  0.01\% &   \\
  b & 6000 &   6015 &   0.08\%   &   0.33\% &   1.68\%  &  0.02\% & x \\
  c & 6015 &   6030 &   0.02\%   &   0.78\% &   2.20\%  &  0.00\% & x \\
  d & 6030 &   6050 &   0.29\%   &   0.03\% &   0.35\%  &  0.03\% & x \\
  e & 6050 &   6070 &   0.10\%   &   0.35\% &   1.08\%  &  0.02\% & x \\
  f & 6100 &   6120 &   0.21\%   &   0.65\% &   1.88\%  &  0.01\% & x \\
  g & 6121 &   6140 &   0.30\%   &   2.92\% &   3.89\%  &  0.00\% & x \\
  h & 6143 &   6160 &   0.60\%   &   0.14\% &   1.11\%  &  0.04\% & x \\
  i & 6160 &   6180 &   0.04\%   &   4.09\% &   5.42\%  &  0.00\% & x \\
  l & 6295 &   6305 &   0.11\%   &   0.67\% &   0.78\%  &  0.00\% &   \\
  m & 6311 &   6320 &   0.02\%   &   0.11\% &   1.34\%  &  0.01\% &   \\
  n & 6579 &   6599 &   0.16\%   &   0.09\% &   1.13\%  &  0.04\% &   \\
  o & 6688 &   6702 &   0.00\%   &   0.01\% &   0.22\%  &  0.00\% &   \\
  p & 6703 &   6711 &   0.00\%   &   0.00\% &   0.43\%  &  0.00\% &   \\
  q & 6711 &   6718 &   0.01\%   &   0.09\% &   0.63\%  &  0.00\% &   \\
  r & 7440 &   7470 &   0.16\%   &   0.55\% &   1.16\%  &  0.01\% &   \\
  s & 7697 &   7702 &   0.00\%   &   0.30\% &   0.20\%  &  0.00\% &   \\
  t & 7769 &   7799 &   0.64\%   &   0.36\% &   0.53\%  &  0.68\% &   \\
   \\
  \hline
  \end{tabular}
  \caption[Spectral Segments]{Wavelength ranges of spectral regions used in this analysis.  Those with a check in the VF05 column cover the same wavelengths as segments from \citetalias{2005ApJS..159..141V}.  The letters correspond to those at the tops of the plots in Figure \ref{fig:cum_distros}.  The columns for decreasing \logg\ and \teff\ quantify the total amount of information contributed to the $\chi^2$ determination for each segment as plotted in the figure.}
  \label{table:line_segs}
\end{table*}

In preparing the spectra for analysis we use the same continuum normalization procedure as \citetalias{2005ApJS..159..141V} but also mask out telluric lines as our spectral range now includes regions with significant telluric contamination.  We perform cross correlation with the solar atlas \citep{2011ApJS..195....6W} to find an approximate radial velocity and before shifting the spectrum to observatory wavelengths, we apply a telluric mask based on telluric lines found in the \citet{2011ApJS..195....6W} atlas.  This masking had a beneficial effect on continuum normalization in some segments for spectra where telluric lines fell on continuum features near the ends of the segment.

\subsection{The New Analysis Procedure}
Initial stellar parameters are all set to solar values except for the temperature, determined by V-K or B-V color relation \ \citep{2013ApJ...771...40B}, and gravity which is arbitrarily set to 4.5. In the first step (``Step 1''), we allow the global parameters \teff, \logg, [M/H], and macroturbulence ($v_{\mathrm{mac}}$) to be free along with individual abundances for Ca, Si, and Ti.  These alpha elements represent the largest number of non iron peak lines in the spectral regions we are analyzing. They also have relatively high abundances that can differ greatly from the solar abundance pattern especially for low metallicity stars.  In our initial fits, overall metallicity is strongly correlated with [Fe/H] due to the preponderance of iron lines.  By letting these alpha elements be independent of overall metallicity, we partially account for non-solar abundance patterns in fitting the global parameters.  \citetalias{2005ApJS..159..141V} showed that microturbulence, important in equivalent width analysis, seems not to play a large role in forward modeling and is degenerate with [M/H] so we fixed it to our adopted solar value of 0.85 km/s.  After fitting, we perturb the temperature $\pm 100$K in the resulting model and re-fit, using these new initial parameters to encourage the non-linear least squares solver to converge to a nearby minimum.  The new analysis provides more consistent results with median standard deviations between the three models of 11K, .01 dex, and .005 dex for \teff\, \logg, and [M/H] for the stars in our sample.

In the next step (``Step 2'') we fix the global parameters to the $\chi^2$ weighted average of the three models from Step 1.  We then allow all of our elemental abundances to be free including the three alpha elements which were free in Step 1.  We have a total of 15 free elemental abundances (C, N, O, Na, Mg, Al, Si, Ca, Ti, V, Cr, Mn, Fe, Ni, Y), which were selected based on our interest in the element, the number of lines available in our spectral range, the sensitivity of those lines to small changes in abundances, and the reliability of recovering the solar abundances in our asteroid spectra (see below).  A few of those elements, carbon, nitrogen, oxygen, and magnesium are also components of molecular species with lines in our spectral range and our inclusion of molecules in the spectral synthesis model provide additional constraints on these elemental abundances.  The resultant model now has the full set of global parameters and individual abundances.

We then iterate, repeating Step 1 and Step 2 using the abundance pattern derived from the output of the first iteration instead of solar values. The differences in the models between the first and second iterations are small but serve to erase a subtle trend in $\Delta$ \logg\  as a function of metallicity that was seen at the end of the first iteration. The standard deviation of the differences between the first and second iteration is 0.056 dex in \logg\ and only 0.01 dex for [M/H].  

\begin{figure}[h!] %  figure placement: here, top
   \centering
   \includegraphics[width=0.5\textwidth]{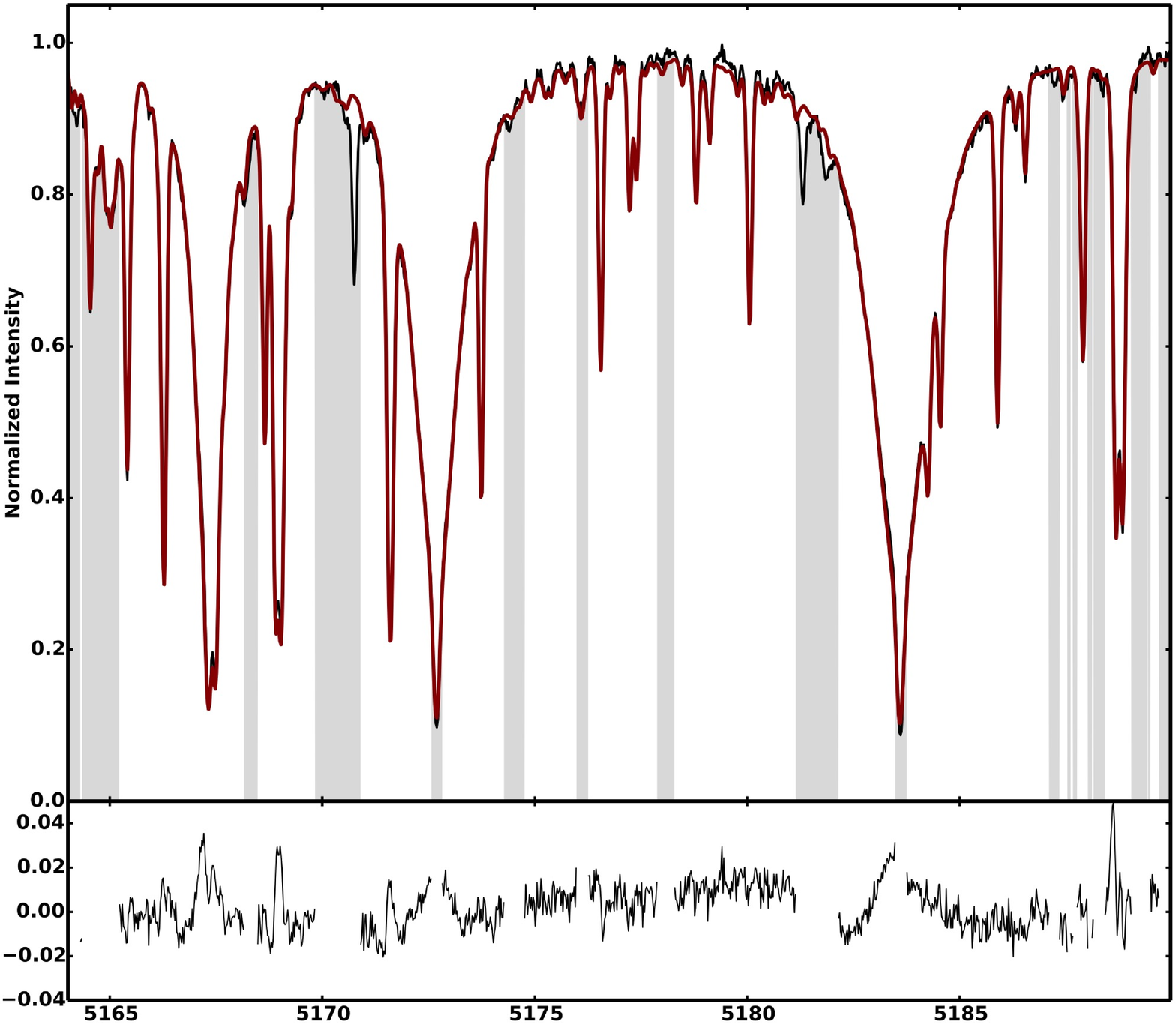} 
   \caption{The wings of the Mg I b triplet contain significant gravity information in our spectral region.  As can be seen in this fit (heavy red line) to a spectrum from the asteroid Vesta (thin black line), we obtain a good fit using our new analysis. Gray regions are excluded from the fit and residuals (observed - model) are shown below.}
   \label{fig:solar_spectrum_fit}
\end{figure}

To evaluate the accuracy of our returned parameters, we analyzed 20 spectra of reflected sunlight from 4 different asteroids taken on 6 different epochs.  These spectra were obtained in the same manner as our other targets.  Because we calibrated the atomic line data to the high resolution NSO solar spectrum, we expect that we should recover the solar parameters when we fit the reflected solar spectra in the asteroid observations.  We found that our procedure recovers the solar parameters for the asteroid spectra (Figure \ref{fig:solar_spectrum_fit}) with  $\chi^2$ weighted mean offsets of only 1K in \teff, 0.02 dex in \logg, and 0.02 dex in [Fe/H] and RMS scatter of 5K, 0.006 dex, and 0.003 dex respectively.

\subsection{Microturbulence}\label{subsec:microturbulence}
	Our analysis is one of differential solar measurements made after tuning our line list against a solar atlas.  In doing so, we adopted a microturbulence value of 0.85 km/s for the sun and left that fixed in the analysis of other stars.  After fitting using our new procedure, we analyzed the effects of this decision to follow \citetalias{2005ApJS..159..141V} in fixing the microturbulence to the value we adopted for the sun with two tests.  The first fixed \vmic\ using the final parameters from our analysis in the empirical formula of \citet{2013ApJ...764...78R}
	\begin{eqnarray} \label{eqn:vmic}
	v_{mic} & = & 1.163  \nonumber \\
	& & + 7.808 \times 10^{-4} \times (T_{\rm eff} - 5800) \nonumber \\
	& & - 0.494 \times (\log g - 4.30) \nonumber \\
	& & - 0.05 \times [Fe/H] 
	\end{eqnarray}
	and then ran our analysis as before.  For the asteroid spectra \vmic\ = 1.07 was 0.2 km/s higher than our value of 0.85 and resulted in models on average 30K hotter and 0.03 dex higher in \logg.  For the asteroseismic stars, the resultant gravities were discrepant from asteroseismic values by up to 0.35 dex and there were clear trends in $\Delta \log g$ with both \teff\ and [M/H]. Temperatures also increased for hotter stars by up to 200K.  Allowing for the possibility that the offset in \vmic\ at solar from our adopted value could be responsible for the differences, we re-ran this test after subtracting a constant 0.22 km/s from the \vmic\ values.  By construction, this then returned the solar parameters for the asteroid spectra but the trend in $\Delta \log g$ with \teff\ remained with a slightly shallower slope ($\Delta \log g$ up to 0.23 dex).  
		
	The second test allowed \vmic\ to be an additional free parameter in the global parameter step with its initial value set to 0.85 km/s and our procedure was then run as before.  There was a systematic offset of $\sim 0.2 km/s$ lower in the returned \vmic\ with respect to the fixed test values.  The asteroid spectra returned the solar parameters with only [M/H] increasing slightly by 0.01 dex.  However, asteroseismic stars again showed clear trends in $\Delta \log g$ with both \teff\ and [M/H] though slightly less ($\Delta \log g$ up to 0.25 dex) than the fixed \vmic\ case.  
	
	The first step in our analysis was to empirically tune our line data against a solar atlas using fixed solar parameters including 0.85 km/s for microturbulence.  Using the formula of \citet{2013ApJ...764...78R} for microturbulence when fitting stellar spectra degraded our gravity accuracy because our lines had been tuned at this constant, lower value.  It is possible that self-consistent use of the \citet{2013ApJ...764...78R} formula to tune the line data and fit stellar spectra might improve the accuracy of our abundance determinations without compromising the accuracy of the gravity.
	
	In our models, microturbulence is partially degenerate with \logg , \teff\ and [M/H].  Including it as an additional free parameter reduces our ability to recover accurate surface gravities.  Solving for microturbulence while tuning the atomic line data might allow us to solve for microturbulence when fitting spectra without compromising the fit for solar-type stars. From these tests it is clear that we have to leave \vmic\ set at the value we adopted for the sun since this is the value used in tuning our lines against the solar atlas.

% ========================================================
% Results Section
%
\section{Results}\label{sec:results}

% 
% SME Classic vs. SME Step: Asteroseismic Sample
%
\begin{figure}[h!] %  figure placement: here, top
   \centering
   \includegraphics[width=0.5\textwidth]{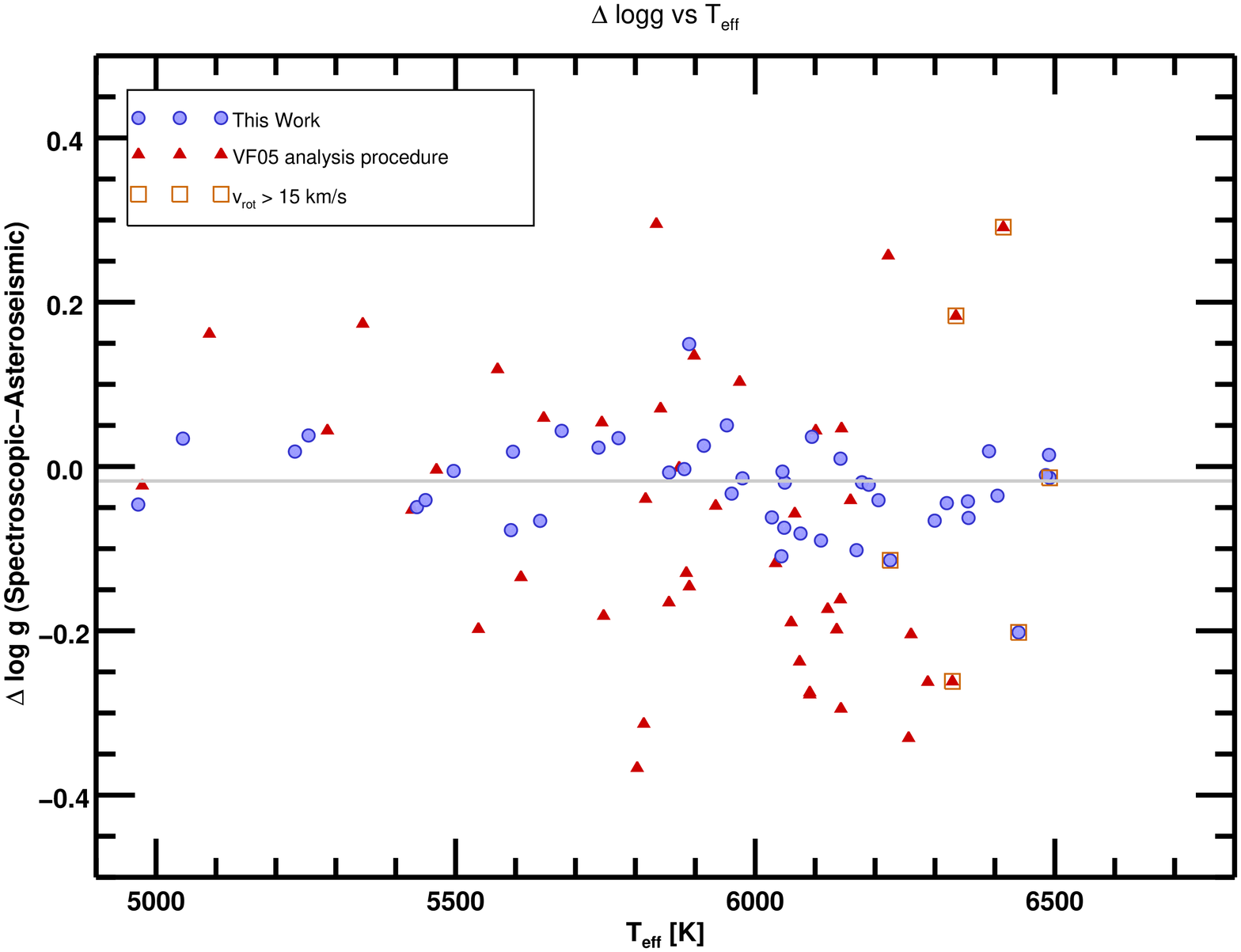} 
   \caption{The unconstrained values of \logg\ determined by our new analysis are much closer to the asteroseismic values than those returned by our analysis using the \citetalias{2005ApJS..159..141V} Line list.  Moreover, there are no trends with temperature and comparisons with gravity and metallicity also show no trends. The lone outlier in our new analysis comes from a star with total rotational broadening ($v_{\mathrm{rot}}$) greater than 25 km/s.}
   \label{fig:sme_v_step}
\end{figure}

Comparing all of the stellar parameters derived from spectral analysis to known stellar parameters is only possible for the Sun, where we have independent methods for determining age and composition.  Even then, debate continues surrounding the accuracy of abundance patterns.  However, for individual parameters such as surface gravity, we now have two methods that provide consistency checks for our analysis.  

\begin{figure*}[h!] %  figure placement: here, top, bottom, or page
   \centering
   \includegraphics[width=\textwidth]{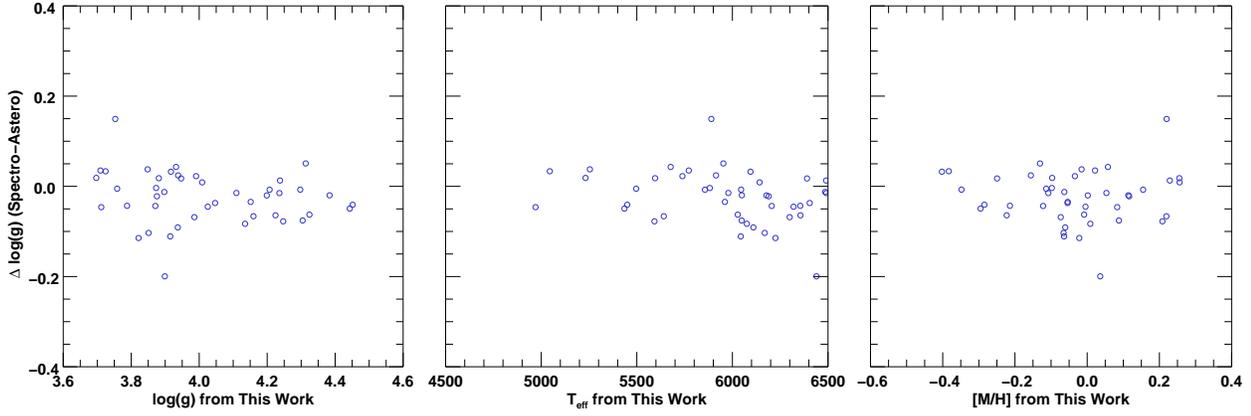} 
   \caption{The spectroscopically determined gravities return the asteroseismically determined gravities with only \seismicRMS\  RMS scatter and show no trends with \logg, \teff, or [M/H]. In the first panel there is no trend in derived gravity for subgiants through solar type dwarfs. In the second panel, we see no trends in derived gravity for stellar temperatures between early K and late F objects.  In the final panel we see no trends for stars with metallicities between -0.5 and  0.3 dex. The outlier at \outlierDlogg\ is a star with total rotational broadening greater than 25 km/s}
   \label{fig:results_asteroseismic}
\end{figure*}

\subsection{Comparison to Asteroseismic Surface Gravity}
The first of these is to compare our spectroscopic gravity with the surface gravity from the grid based asteroseismic method described in \S\ \ref{sec:astero_logg} as our fiducial.  Our new analysis process results in surface gravities which match the asteroseismically determined values of the stars in our sample with an RMS scatter of \seismicRMS\  and with virtually no offset (\seismicOffset) in the zero-point.  Improvements with respect to the prior analysis using the \citetalias{2005ApJS..159..141V} procedure can be seen in Figure \ref{fig:sme_v_step}.  There are no significant trends in $\Delta$ \logg\ between the asteroseimically and spectroscopically determined values with respect to \teff, \logg, or metallicity (Figure \ref{fig:results_asteroseismic}).  Errors in these three parameters have been shown to be correlated in previous spectral synthesis modeling analyses using the \citetalias{2005ApJS..159..141V} line list and an older SME version \citep{2012ApJ...757..161T}. The extreme outlier at $\Delta$ \logg\ $=$ \outlierDlogg\ (Figure \ref{fig:results_asteroseismic}) is a star with total rotational broadening ($v \sin i$ and $v_{\mathrm{mac}}$) greater than 25 km/s which smooths away most of the information needed to determine gravity and temperature.

% ========================================================
% Comparisons
%
\subsection{Comparison with Published Results}\label{sec:comparison_published}

\begin{figure*}[h!] %  figure placement: here, top, bottom, or page
   \centering
   \includegraphics[width=\textwidth]{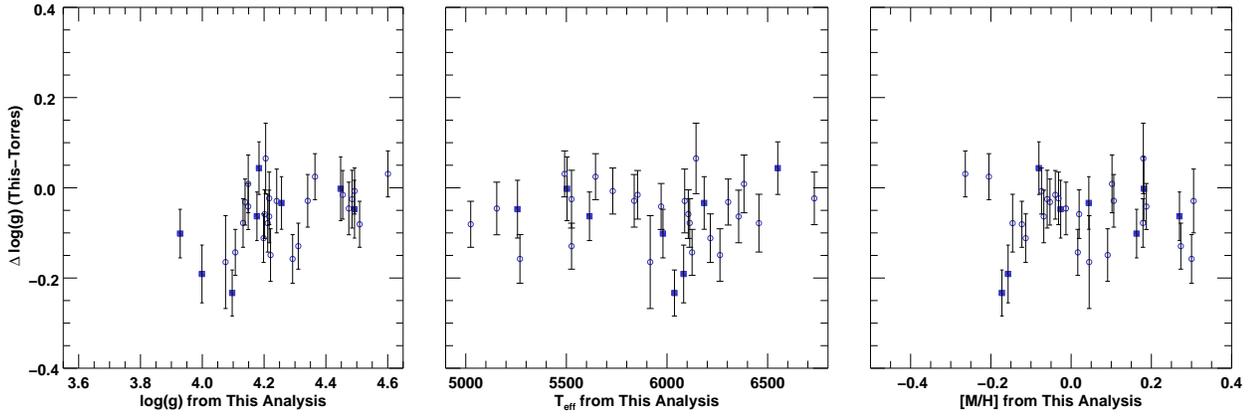} 
   \caption{A comparison of stars analyzed by \citet{2012ApJ...757..161T} made use of $a/R_*$ for transiting planets. There does not appear to be any trend in derived surface gravity with respect to the parameters \logg, \teff\, or [Fe/H].  This sample contains several low mass stars missing in the asteroseismology sample. Spectra with SNR $< 100$ per wavelength bin have square markers.}
   \label{fig:results_torres}
\end{figure*}

A second comparison of our spectroscopic analysis comes from the gravity constraint of transiting exoplanets.  We compared our spectroscopic \logg\ values with those derived by \citet{2012ApJ...757..161T} using the $a/R_*$ density method \citep{Sozzetti:2007ej} for stars with transiting planets. Most of the spectra have SNR$ > 100$ per wavelength bin, though a quarter of them have SNR below 100.  \citet{2012ApJ...757..161T} found systematic trends between their gravities based on $a/R_*$ and SME gravities obtained using the \citetalias{2005ApJS..159..141V} procedure.  In contrast, we find no systematic trends between \citet{2012ApJ...757..161T} gravities and our SME gravities obtained using more spectral segments and a two-stage fitting procedure.  Our SME methodology works better than the  \citetalias{2005ApJS..159..141V} procedure for the \citet{2012ApJ...757..161T} stars, which are generally warmer than the original \citetalias{2005ApJS..159..141V} sample.  We find a constant offset of \torresOffset\ in \logg, relative to \citet{2012ApJ...757..161T}.  Such a small offset could be a result of errors in our spectroscopic analysis or may be the result of small errors in the $a/R_*$ analysis that arise because of inaccuracies in the impact parameters or eccentricities \citep{2013ApJ...767..127H}.

% ========================================================
% Discussion
%
\section{Discussion}\label{sec:discussion}

We believe that the largest improvement in our determination of surface gravity was a result of adding new wavelength segments.  The expanded line list added new gravity and temperature dependent lines that helped reduce parameter degeneracy.  We also included a factor of 10 more iron lines in the mask, bringing the total to $\sim 900$ with almost one third of them \ion{Fe}{2} lines.  All of the iron lines, regardless of ionization state or excitation potential, are modeled with the same iron abundance.  Because most of the iron is in an ionized state at the temperatures we are interested in, weak \ion{Fe}{2} lines will become stronger with decreasing gravity while weak \ion{Fe}{1} lines are largely gravity insensitive.  Having a large number of lines in both states provides another important constraint on the gravity and helps to further limit the parameter space.  The addition of the expanded wavelength range gave us very accurate gravities and left us with only a small residual trend with metallicity.

\begin{figure}[!h] %  figure placement: here, top
   \centering
   \includegraphics[width=0.48\textwidth]{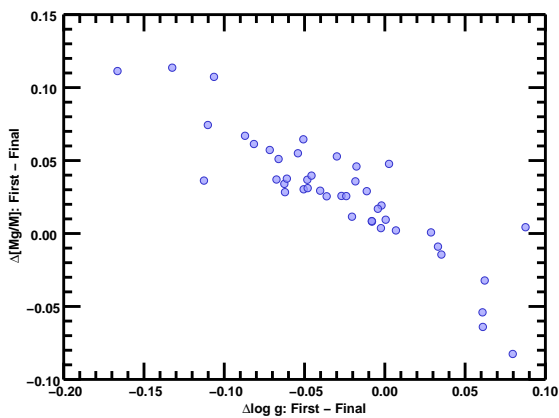} 
   \caption{During the first iteration of our analysis, the Mg abundance is just the solar abundance pattern scaled to the overall metallicity.  After obtaining a close estimate of the global parameters we get more detailed abundances for 15 elements including Mg and this new abundance pattern is used in fitting for the final global parameters.  Because Mg lines constitute the bulk of the surface gravity information for cool dwarfs, this iterative process allowed us to remove subtle trends in derived $\log(g)$ with metallicity in our sample.}
   \label{fig:mgm_v_logg}
\end{figure}

The wings of the Mg I b triplet lines provide a wealth of information on the surface gravity of a cool dwarf star, but the gravity information is degenerate with the Mg abundance. This led us to separate the global parameter fitting from the abundance determinations.  However, because the particular atmosphere used is tied to the metallicity, a scaled solar abundance pattern (dominated by [Fe/H]) will not work for all stars.  By allowing the most abundant alpha elements (minus Mg) to be independent of the overall metallicity we obtained a closer match to the correct atmosphere on the first iteration.  The magnesium abundance, however, will be the scaled solar value in this step and not necessarily accurate.  In the second iteration, we have a direct determination of the magnesium abundance and so obtain more accurate gravity information from the Mg I b wings. In Figure \ref{fig:mgm_v_logg} we see that the changes in the magnesium to metallicity ratio between the first step (scaled solar value) and the final (independently determined Mg abundance) are inversely proportional to the change in gravity.

As K and M dwarfs are some of the most promising candidates for finding Earth-like planets in the habitable zone a robust spectroscopic analysis of cool dwarfs is an important improvement to the work presented here.  We are currently finishing the analysis of all stars observed by the California Planet Search since late 2004 using this procedure and will include our updated line list in that paper.

% ========================================================
% Conclusions
%
\section{Conclusions}\label{sec:conclusions}

Accurate stellar parameters can be obtained from high resolution spectroscopic analysis when spectral segments contain adequate constraints and care is taken to decouple degeneracies in the fitting process.  The results improve upon our past analyses where correlated errors led to surface gravity determination off by more than 0.3 dex for stars where Mg I b is no longer a good gravity constraint.  Our technique returns gravities consistent with those determined by asteroseismic analysis with an RMS scatter of only \seismicRMS.  It is also consistent with gravities determined using $a/R_*$ constraints from planetary transits with a small systematic offset of \torresOffset.

% ========================================================
% Acknowledgements
%
\acknowledgments
This work was supported by a NASA Keck PI Data Award, administered by the NASA Exoplanet Science Institute. The authors thank the anonymous referee for their insightful comments which greatly improved the quality of this work. Data presented herein were obtained at the W. M. Keck Observatory from telescope time allocated to the National Aeronautics and Space Administration through the agency's scientific partnership with the California Institute of Technology and the University of California. The Observatory was made possible by the generous financial support of the W. M. Keck Foundation.

The authors wish to recognize and acknowledge the very significant cultural role and reverence that the summit of Mauna Kea has always had within the indigenous Hawaiian community. We are most fortunate to have the opportunity to conduct observations from this mountain.

DF and JMB acknowledged  NASA grant NNX12AC01G. SB acknowledges support from NSF grant AST-1105930 which also partially supported JMB. SB was further supported by NASA grant NNX13AE70G.
% ========================================================
% References
%
\bibliography{ms}

\end{document}